# Density Functional Theory Analysis that Explains the Volume Expansion in Prelithiated Silicon Nanowires

Donald C. Boone
Nanoscience Research Institute, Arlington, VA 22314 USA; db2585@caa.columbia.edu

**Abstract:** This research is a theoretical study that simulates the volume expansion of a prelithiated silicon nanowire during lithium ion insertion and the application of an electric current. Utilizing density functional theory (DFT) the ground state energy $E_g(x)$ of prelithiated silicon ($Li_xSi$) is define as a function of the lithium ion (Li+) concentration (x). As the Li+ are increased, $E_g(x)$ becomes increasingly stable from x=1.00 through x=2.415 and decreases in stability as the lithium ion concentration becomes x >2.415 until full lithiation of the silicon nanowire is reached at x=3.75. After the determination of the lithiated silicon ground state energies an electric current is applied to the lithiated silicon nanowire at varies Li+ concentrations x. It was discovered that the volume expansion began at approximately x=3.25 and increased to over 300% of the original volume of a pristine silicon nanowire at x=3.75 which at this point was full lithiation. This is in sharp contrast to prior research studies where the ground state energy was not considered. In previous studies the computation of the volume expansion starts approximately at x=0.75 and produces a continuous nonlinear volume expansion until the process was terminated at full lithiation.

## Introduction

Concurrently as lithiated silicon was being studied, varies research groups were developing the process of prelithiation [1]. The main reason to prelithiate a silicon nanowire is to prevent lithium loss that can occur in three different ways: 1) To prevent the formation of the solid electrolyte interphase (SEI) layer that consumes lithium ions, 2) after lithiation a minority of lithium ions are unable to delithiate due to the formation of highly stable lithiated silicon molecules at defect sites. A large amount of lithium ions are stored on these defects which cannot be released again resulting in lithium ion losses and 3) lithium ions are trapped within interstitial sites in the silicon anode that accounts for approximately 30% of the initial lithium ion loss during the first lithiation cycle [2]. This leads to accelerated decay of the silicon anode material in the subsequent cycles during volumetric expansion [3].

One approach to fabricate prelithiated silicon nanowires is to embed lithium ions in the silicon anode which would replenish the lithium ion losses during lithiation cycles. As a result of this process severe fracture and pulverization of Si-based particles and excessive side reactions between silicon anode and lithium ions would be avoided, resulting in improved energy density of lithiated silicon nanowires [4].

## Theoretical Framework

The solution of the $Li_xSi$ ground state energies $E_g(x)$ will be performed using the self-consistent Kohn-Sham equation of density functional theory (DFT). The Hartree-Fock approximation is used to construct the Hamiltonian H where the exchange-correlation functional is present [5]. The computational model for this study was constructed using multiple silicon cubic diamond lattices that are simulated to be present within the silicon nanowire.

## Ground State Energy

The computational results of the prelithiated silicon ground state energies $E_g(x)$ are of a parabolic curvature form due to the determination of the DFT calculations which are displayed in figure 1. The lithium ion concentration (x) represents the lithium/silicon ratio composition in an individual silicon nanowire. The beginning of the curve at x=1.00 approximately represents the covalent bond energy between two silicon atoms at $E_g(x)$= -8.41 eV. The lithiated silicon concentration at x=1.00 is one lithium ion to one silicon atom ratio and therefore the coulombic energy of the lithium ions are relatively low.

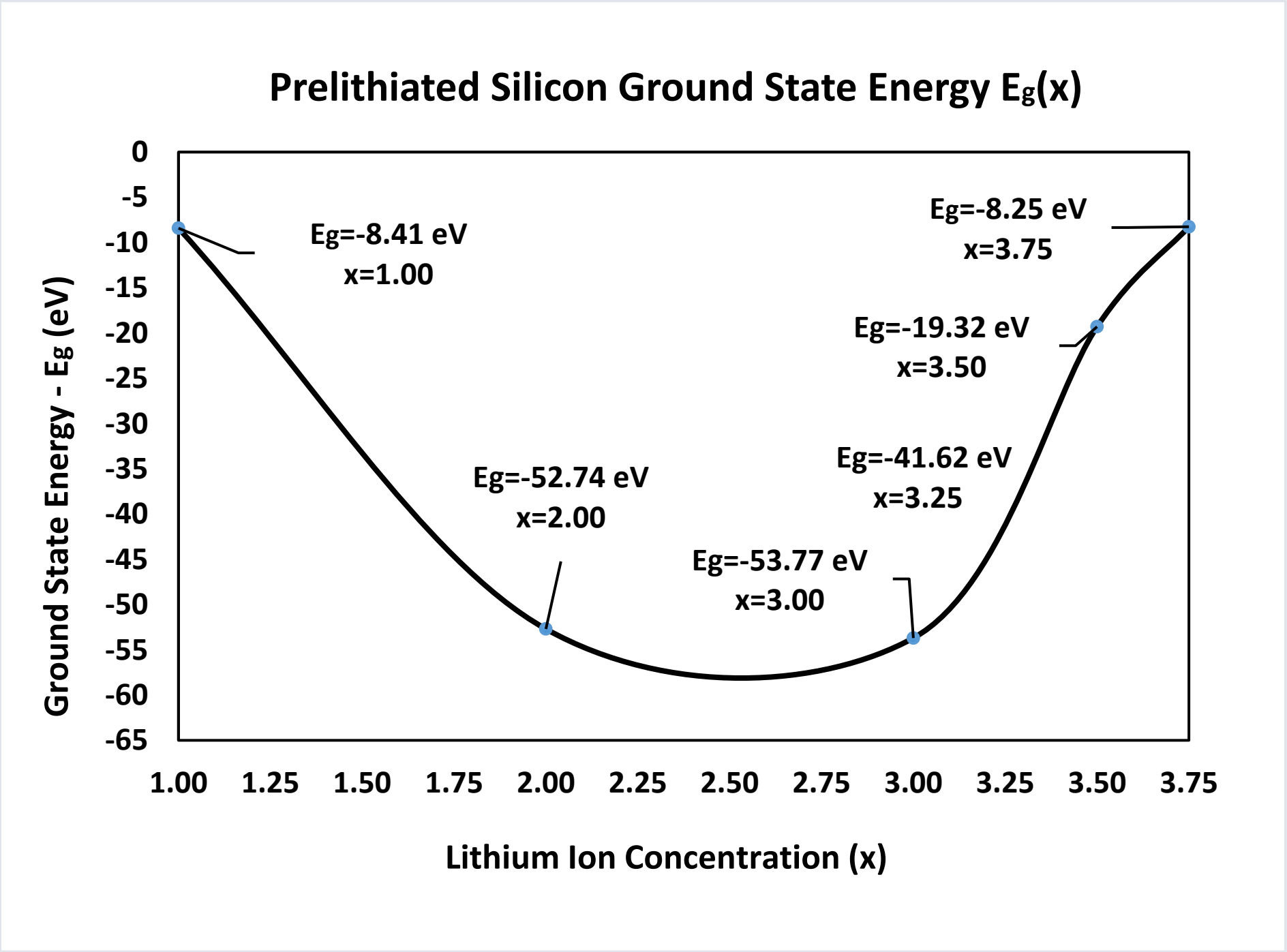

**Figure 1.** The ground state energy decreases and becomes less stable after x=2.415. Starting at x=3.25 the ground state energy $E_g(x)$= -41.62 eV and at full lithiation at x=3.75 $E_g(x)$= -8.25 eV.

As the lithium ion concentration increases from x=1.00 to x=2.40 the ground state energy becomes more stable and at x=2.415, $E_g(x)$= -61.8 eV. This is the lowest and most stable ground state energy. With increasing Li+ concentration the $E_g(x)$ become less stable above x=2.415 when finally at full lithiation at x=3.75, $E_g(x)$= -8.25 eV. In order to better understand the parabolic shape of the prelithiated ground state energy, decomposition is performed on the Hamiltonian in order to separate the energy system into its attractive and repulsive energies in which the sum is known as the coulombic energy as shown in figure 2. The coulombic energy is composed of lithium ions and bonded state electrons from silicon atoms and lithium ions. When the lithium ion concentration is between x=0.25 to x=2.40 the attractive energy (which are the interactions between positive lithium ions and negative bonded state electrons) within silicon atoms and lithium ions is vastly predominate over the repulsive energy which is defined as the interactions between positive lithium ions. However beyond x=2.40 the repulsive energy experience a linear exponential increase until full lithiation at x=3.75. This causes the sudden tendency of the coulombic energy at approximately x=2.40 to seek a balance between the attractive and repulsive energies and at x=3.25 the two opposing energies become equal and opposite in magnitude and the normalized coulombic energy becomes zero. Between x=0.25 and x=3.25 the coulombic energy is more attractive energetically, however beyond x=3.25 the coulombic energy becomes exponentially repulsive.

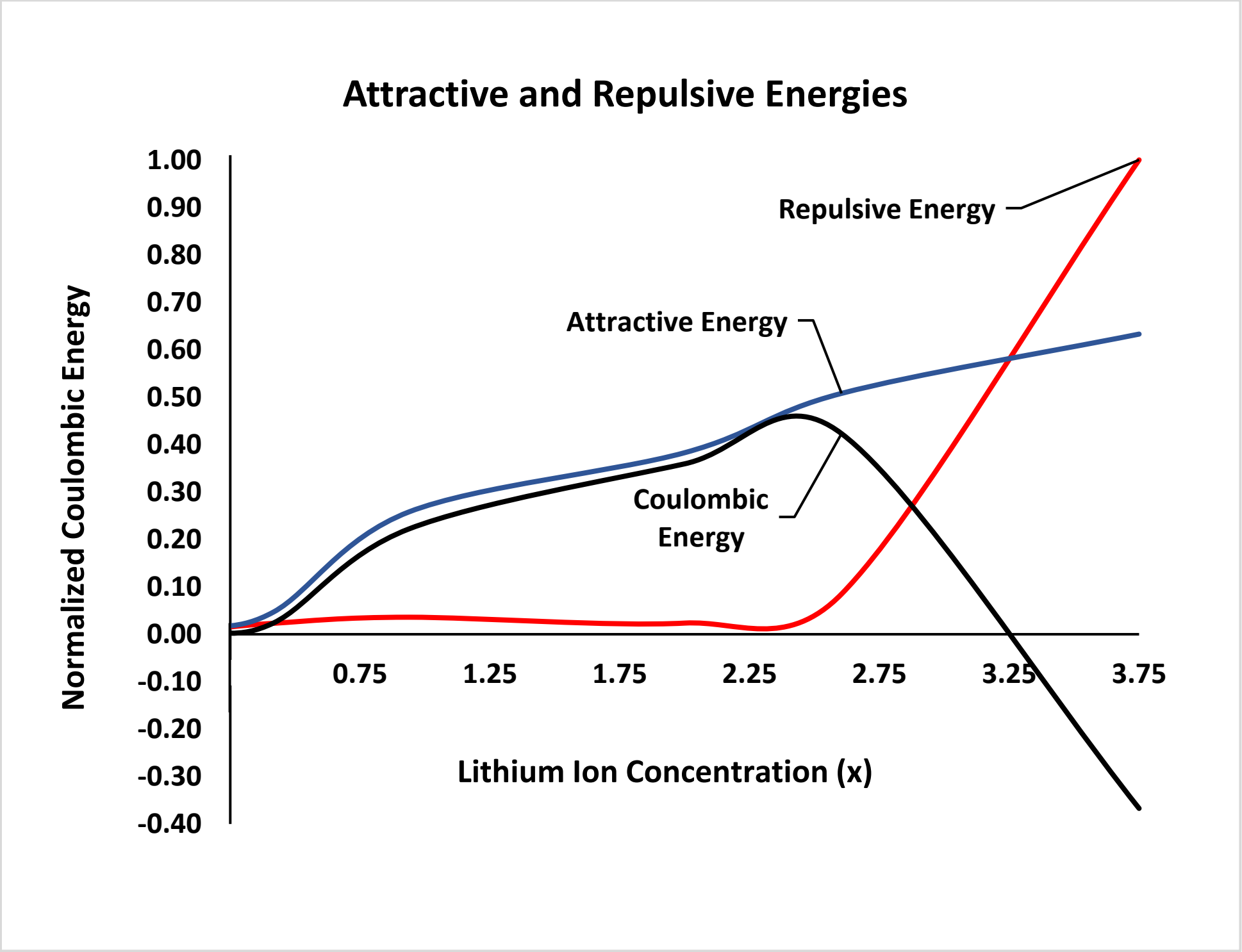

**Figure 2.** From x=0.25 to x=3.25 the coulombic energy is more energetically attractive (normalized positively), however beyond x=3.25 the coulombic energy becomes energetically repulsive (normalized negatively).

## Electric Current

A computational simulation of an electric current driven by a 2-voltage source is applied to a prelithiated silicon nanowire during lithium ion insertion to calculate the volume expansion. This research was previously performed on a silicon nanowire that was not prelithiated in which lithium ions were diffused through one end of a silicon nanowire (and therefore becoming lithiated) with an electric current traveling through the opposite end of the nanowire [6].

The coherent energy $E_c(x)$ from the electric current is derived from photons in which there frequencies are in-phase with similar frequencies and capable of producing the require energy to excite the prelithiated silicon molecular particles from the ground state. The photons are modeled after a cumulative distribution function of a Poisson distribution to calculate the probability that a sufficient coherent energy is being generated to produce a volume increase in the lithiated silicon nanowire. In figure 3 there are three cumulative distribution functions that are displayed for three separate lithium ion concentrations (x) for three individual prelithiated silicon nanowires. In order for volume expansion of the nanowires to occur the ground state energy of the prelithiated silicon must be perturbed in which excitation of the bonded electrons is raised to higher energetic state to an extent where the bonds between the lithium and silicon particles are severed. When x=3.75 the coherent energy is $E_c(x)$=19.72 eV which correspond to a ground state energy $E_g(x)$= 8.25 eV at full lithiation. The probability that bonded state electrons of prelithiated silicon material will enter into an excited state is at 0.99985. Since this probability is approaching one, it is extreme likely that the electrons will reach excitation and the bonds of the constitutive particles will break and volume expansion of Vc=303.54 will develop. For a prelithiated silicon nanowire of x=3.50 the ground state energy is $E_g(x)$=19.32 eV while the coherent energy is $E_c(x)$ =15.19 eV. This will correspond to a probability of 0.19590 that molecular prelithiated silicon will be excited in order for volume expansion Vc=151.78 to occur. When the prelithiation of the silicon nanowire is at x=3.25 the ground state energy is $E_g(x)$ =-41.60 eV with a coherent energy $E_c(x)$ =11.81 eV. The probability that the volumes will increase due to severing of molecular bonds is extremely low at P=0.0007 however this does corresponds to a volume change of Vc=20.34. At lower lithium ion concentrations at x< 3.25 the coherent energy is too low to break molecular bonding and therefore volumetric expansion approaches zero and is negligible.

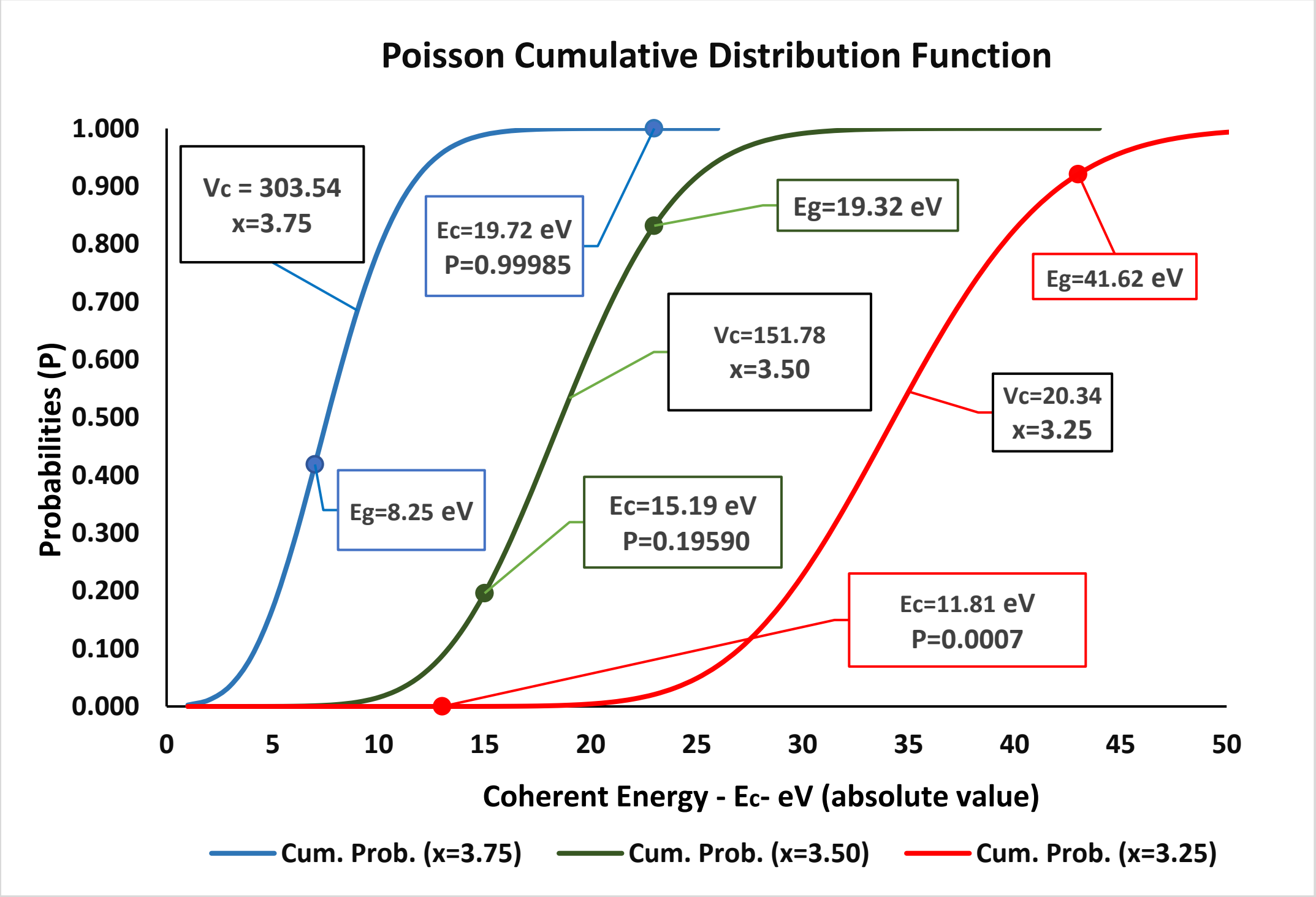

**Figure 3.** The ground state energy $E_g(x)$ and coherent energy $E_c(x)$ are both dependent on the lithium ion concentration (x). Each cumulative distribution function represents a prelithiated silicon nanowire with a specific (x) value. At x=3.75 the $E_g(x)$=8.25 eV corresponds to a $E_c(x)$ = 19.72 eV. Since the coherent energy, which is a collection of photons, is larger in energy than the ground state energy, the probability P=0.99985 that the lithium-silicon or silicon covalent bonds will severe is almost guaranteed. As the lithium ion concentration (x) becomes lower for any particular prelithiated silicon nanowire the coherent energy $E_c(x)$ becomes lower which results in a probability that the molecular bonds will not break and therefore volumetric expansion will not occur.

## Volumetric Expansion

As previously mentioned, there was a former research study that simulated an applied electric current to a silicon nanowire in order to calculate the volume expansion during lithium ion insertion [7,8]. The results from that prior research is displayed in figure 4 which is labeled 'no prelithiation' along with the research results that is labeled 'prelithiation' which is based on the density function theory study that is presented in this report. The volume expansion of the silicon nanowire that was not prelithiated has a constant volume change from x=0.75 to approximately x=1.25 and at x>1.25 the volume change increases at a nonlinear rate until full lithiation is achieved at x=3.75 at which point the lithiated silicon nanowire volume change is approximately 300%. However the prelithiated silicon nanowire has a negligible volume change from x=0.75 to x=3.25. At the lithium ion concentration of x=3.25 a volume increase of 20.34% occurs and continues to increase linearly until full lithiation at x= 3.75 at which point the volume change will be approximately 300% of the original volume of the silicon nanowire.

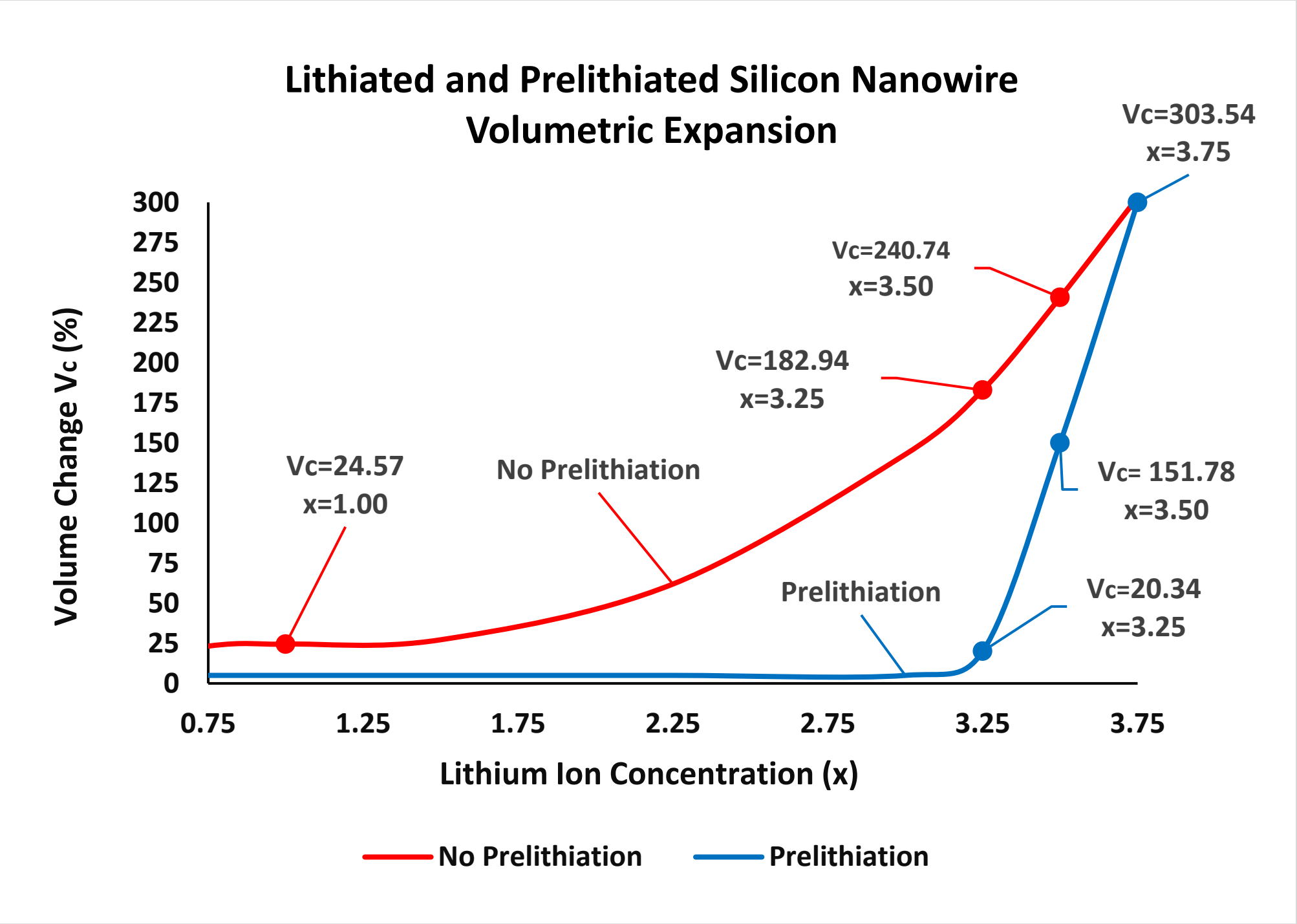

**Figure 4.** This data demonstrates that the prelithiated silicon nanowire volume increase during the first lithiation cycle is near zero when the lithium ion concentration (x) for the silicon nanowire is between x=0.75 to x=3.25. The silicon nanowire that has not been prelithiated experiences a nonlinear volumetric expansion increase at any specific (x) value.

## <u>Conclusion</u>

This research study demonstrates through the use of density function theory that the fabrication process of prelithiation can be a viable solution to control the volume expansion of lithiated silicon nanowires under certain lithium ion concentrations (x). From x=0.25 to x=3.25 the prelithiated silicon volume expansion is near zero. In comparison to the lithiated silicon nanowire that has not been through the prelithiation process the volume increase between x=0.25 to x=3.75 is approximately 25% to 300% respectively. However, prelithiation between x=3.25 through x=3.75 can develop similar volume expansion problems as the lithiated silicon nanowires. Therefore, great care should be performed in the application of the lithium ions insertion during prelithiation process in order to avoid the volume increase within the silicon nanowires.

## <u>Conflict of Interest</u>

The author declare no conflict of interest.